\documentclass[11pt,letter]{article} 
\usepackage{jheppub} %

\usepackage[ps,dvips,matrix,arrow,frame,import,curve,color]{xy}

\notoctrue

\def\blfootnote{\xdef\@thefnmark{}\@footnotetext}

\long\def\symbolfootnote[#1]#2{\begingroup%
\def\thefootnote{\fnsymbol{footnote}}\footnote[#1]{#2}\endgroup}

\newcommand{\be}{\begin{eqnarray}}
\newcommand{\ee}{\end{eqnarray}}
\newcommand{\ben}{\begin{eqnarray*}}
\newcommand{\een}{\end{eqnarray*}}

\newcommand{\bcent}{\begin{center}}
\newcommand{\ecent}{\end{center}}
\newcommand{\benum}{\begin{enumerate}}
\newcommand{\eenum}{\end{enumerate}}
\newcommand{\bdesc}{\begin{description}}
\newcommand{\edesc}{\end{description}}
\newcommand{\bitem}{\begin{itemize}}
\newcommand{\eitem}{\end{itemize}}
\newcommand{\bquote}{\begin{quote}}
\newcommand{\equote}{\end{quote}}
\newcommand{\bhalfp}{\begin{minipage}{0.45\textwidth}}
\newcommand{\ehalfp}{\end{minipage}}
\newcommand{\bhead}{\begin{center}\bf \Large}
\newcommand{\ehead}{\end{center}\bigskip}

\def\be{\begin{equation}}
\def\ee{\end{equation}}
\def\ba{\begin{eqnarray}}
\def\ea{\end{eqnarray}}

\newcommand{\roughly}[1]{\mathrel{\raise.3ex\hbox{$#1$\kern-0.85em
\lower1ex\hbox{$\sim$}}}}

\def\2pi{\left(2\pi\right)}

\def\beq{\begin{equation}}
\def\eeq{\end{equation}}
\def\bg{\begin{eqnarray}}
\def\nd{\end{eqnarray}}
\def\bea{\begin{eqnarray}}
\def\eea{\end{eqnarray}}

\def\D3{\overline{\mbox{D3}}}

\title{A Note on the Stringy Embeddings of Certain ${\cal N} = 2$ Dualities}

\author{Keshav Dasgupta}
\author{and Jihye Seo}
\affiliation{Ernest Rutherford Physics Building, McGill University,\\
3600 University Street, Montr{\'e}al QC, Canada H3A 2T8}
\emailAdd{keshav, jihyeseo@hep.physics.mcgill.ca}
\abstract{Seiberg-Witten theory can be embedded in F-theory using D3 branes probing an orientifold geometry. The non-perturbative corrections in the 
orientifold picture map directly to the instanton corrections in the corresponding gauge theory that convert the classical moduli space to the quantum one. In this
short review we argue that the recently proposed class of conformal Gaiotto models may also be embedded in F-theory. The F-theory constructions help us 
not only to understand 
the Gaiotto dualities but also 
to extend to the non-conformal cases with and without cascading behaviors. For the conformal cases, the near horizon
geometries in F-theory capture both the UV and IR behaviors succinctly.} 
\begin{document}
\maketitle
\flushbottom

\newpage

\section{Introduction}

The exact solution of a class of low energy ${\cal N} = 2$ gauge theories with at most two derivatives and four fermion couplings by Seiberg and Witten \cite{sw1, sw2} 
has received wide applications not only in physics but also in mathematics. Interestingly, with every such applications, new aspects of the theory 
come forth that further enrich our understanding of ${\cal N} = 2$ theories.  One such recent work is by Gaiotto \cite{gaiotto} who analyzes Seiberg-Witten theory using
wrapped five-branes on Riemann surfaces in M-theory. 
Although this point of view is not completely new $-$ the original work being done by Witten \cite{wittenM} $-$ the final 
results turned out be a novel class of ${\cal N} = 2$ {\it conformal}
gauge theories that were not considered
in \cite{sw1,sw2}. Interestingly a hint that there would exist such a 
novel class of ${\cal N} = 2$ gauge theories came from an earlier paper by Argyres and Seiberg \cite{sa} done almost two years before Gaiotto's work \cite{gaiotto}.     

The unifying theme between the original Seiberg-Witten theory \cite{sw1,sw2} and the recent Gaiotto theories \cite{sa, gaiotto} is the underlying M-theory construction 
that we discussed above. All these theories come out from M-theory wrapped branes. However one may also {\it derive} most of the properties of the original 
Seiberg-Witten theory using branes in F-theory \cite{senF, bds}. A natural question then is to inquire whether we
can derive the new class of Gaiotto models from F-theory. The answer 
turns out to be in affirmative as shown in \cite{DSW}. In the following we will briefly elucidate this picture and show that the F-theory model of \cite{DSW} is 
powerful enough to extend the original conformal constructions of Gaiotto to non-conformal and cascading ${\cal N} = 2$ gauge theories. 

\section{F-theory embeddings of ${\cal N} = 2$ gauge theories}

The models that we want to study here
all belong to the same class of multiple D3-branes probing the F-theory backgrounds of
seven-branes and orientifold seven-planes. 
The simplest
configuration is then $r$ D3-branes probing the usual parallel D7/O7 background along flat directions, 
giving $Sp(2r)$ Seiberg-Witten theory with antisymmetric matter. We can compare the brane picture and 
SW geometry. 
As a first step toward proving F-theory/SW-geometry duality for higher ranks, we can start by studying the pure $Sp(2r)$ SW curve, for the time being ignoring all the matter effects, analyzing the moduli spaces of both the
theories. 
A non-trivial yet manageable step would be 
to analyze some special points on the moduli space rather than the full space, such as the maximal Argyres-Douglas points as 
we showed in \cite{DSW}. 
From the above two analysis, one may now 
discuss a potential 1-1 map between the moduli space of the brane dynamics and that of the Seiberg-Witten curve for higher ranks. 
At rank 1, antisymmetric matter is empty, so the mapping is rather simple.
However for higher ranks we cannot ignore the anti-symmetric matter and therefore
it does make a difference.

In fact this is where we face some non-trivial issues. 
The 1-1 mapping becomes rather
subtle even in the limit where we take the antisymmetric matter to be heavy, 
and in \cite {DSW} we point out various issues related to this, speculating on some ways to resolve the puzzles. There may exist an 
ideal setting where such 1-1 mapping could in-principle be realized, as we elucidate this in the appendix of \cite{DSW}.

With the same aim in mind, singularity structure of moduli space of pure SW theory with classical gauge groups are studied in \cite{SD}, although corresponding F-theoretic description is not known for most of them. Before one attempts to map the whole moduli space to a dual theory/description, one can start by answering a less ambitious question of how a certain subspace, such as the singularity locus, maps under the duality. 
In \cite{SD}, BPS charges of massless dyons are computed for pure SW theory with $Sp(2r)$ and $SU(r+1)$ gauge groups, and maximal Argyres-Douglas points are computed for pure $SO(2r+1)$ SW curve as well. Like $Sp(2r)$ case of \cite{DSW}, the number of maximal Argyres-Douglas points of $SO(2r+1)$ equals to the dual Coxeter number of the gauge group \cite{SD}. A systematic method of locating Argyres-Douglas locus in pure SW theory is given, and we learn that they occur generically in complex codimension two locus of moduli space. With abundance of Argyres-Douglas theories within SW theory, it will be interesting to have a systematic approach in the spirit of \cite{GST}. Not only that, a description 
in terms of D3-branes probing F-theory geometries would be a welcome advancement in our understanding of the moduli spaces of these theories. 
Though verification and explicit construction needs to come, it is natural to expect that
all the classical gauge groups come from suitable Higgsing of a bigger $Sp(2r)$ theory on the D3-branes probing the simplest F-theory geometry. For example, recall that one obtains $SU(r)$ rather than $Sp(2r)$ as $r$ D3-branes move far away from the orientifold seven-plane.  

Interestingly however, the simplest model 
of D3-branes probing 7-brane geometry
turns out {\it not} to be the most generic one. There is a way to generalize this further {\it without} breaking any more supersymmetries.
This is done by taking the F-theory configuration discussed above and wrapping the
seven-branes and orientifold planes on multi Taub-NUT spaces, which is still ${\cal N}=2$ theory despite the new complication in the geometry. 
With D3-branes probing this background and the seven-branes arranged at the constant coupling points of \cite{DM1},
we argued in \cite{DSW} that these configurations may succinctly explain many of the recently proposed
Gaiotto-type models \cite{gaiotto} on the
world-volumes of the D3-branes. The simplest way to see this would be to decompose the
D3-branes as {\it fractional} D3-branes. These fractional D3-branes are D5-brane anti D5-brane (${\rm D5}$-$\overline{\rm D5}$) pairs, and they
wrap 2-cycles of the multi Taub-NUT (TN) spaces. As shown in details in \cite{DSW},
various ways of wrapping the TN 2-cycles will lead to
various different sets of product gauge groups.  

The wrapped seven-branes on the TN lead to interesting physics when we switch on {\it time-dependent} gauge
fields on the 2-cycles.  These time-varying gauge fields may lead to chiral anomalies along the TN circle (at
infinity). Additionally, the wrapped ${\rm D5}$-$\overline{\rm D5}$ pairs undergo certain interesting transmutations from the
varying vector fields. These transmutations also effect the tachyons between the ${\rm D5}$-$\overline{\rm D5}$ brane pairs.
The Gaiotto dualities (or at least a class of them)
are then explained by three simultaneous effects: chiral anomaly cancellations, anti-GSO projections and brane transmutations.
Following these steps of operations, we elaborated the dualities in the   
various conformal cases of the Gaiotto models in \cite{DSW} reproducing, among others, aspects of the original Argyres-Seiberg duality \cite{sa}. 
Interestingly,
under some special situations, we showed that our model may be dualized to the
brane networks of \cite{bbt} which itself captures a large class of the Gaiotto dualities. These mappings serve as consistency checks of our scenario.

The F-theory configuration with fractional branes and Taub-NUT spaces, turns out to yield other bonuses exposing the power of this approach.  
The first one is the associated gravity duals of the Gaiotto's models in the conformal cases where one may simply consider the metric of the 
wrapped brane configurations and look for the near horizon geometries. 
Since the UV limits are
described by the probe D3-branes
decomposed into ${\rm D5}$-$\overline{\rm D5}$ pairs wrapped on vanishing 2-cycles of multi Taub-NUT space in our models, the
corresponding gravity duals should capture the
six-dimensional gauge theories with the gravity duals having the usual near-horizon $AdS_7$ factors. On the other hand, since
the IR remains a
four-dimensional theory, the gravity duals should again capture the underlying physics with the near horizon geometries having the 
standard $AdS_5$ factors. 
In \cite {DSW} we gave brief derivations in both of
these limits and argued therein how to see the two different scenarios at the UV and IR. It will be interesting to compare these results with \cite{GM}. 

The second unexpected bonus from our F-theory constructions turned out to be the extension to non-conformal scenarios with and without cascading dynamics.  
The cascading models come out from wrapping {\it different} number of D5-branes on the Taub-NUT cycles. On the other hand, the non-cascading 
models appear from rearranging the flavor seven-branes away from the constant coupling scenarios of \cite{DM1} in the limit where the number of colors exceed the 
number of flavors. 
All these 
constructions lead to hitherto unknown new non-conformal models. 
Interestingly, it was suspected for sometime that the non-conformal deformations
of the Gaiotto models should show cascading behaviors. In \cite{DSW} we argued not only why they are most natural, but also showed that there are a class of non-conformal
extensions that do not show cascading behaviors. Recently in \cite{alishaK} the cascading dynamics in these class of models have been discussed in details using 
type IIA brane constructions. 
One may also note in passing that the 
cascading behaviors are different from the standard ${\cal N} = 1$ cascading. These models show ${\cal N} = 2$ cascades \cite{poln2, benini3}. 
These two cascading behaviors could 
in principle be connected, as we showed in \cite{DSW}.

\section{Conclusions}

In this short note we discussed how a slight modification of the standard F-theory configuration with D3 and seven-branes can capture the newly constructed 
Gaiotto models, both for the conformal as well as for the non-conformal cases. The large class of conformal Gaiotto models then appear from wrapping 
${\rm D5}$-$\overline{\rm D5}$ pairs on the two-cycles of multi Taub-NUT spaces, with the seven-branes arranged in the constant coupling scenarios similar to 
\cite{DM1}. The F-theory construction can have further generalisations if we replace the Taub-NUT space with a K3 manifold, or if we have intersecting 
seven-branes. These have been discussed in \cite{DSW}. For the model at hand, many questions still remain. For example how to see the surface operators, 
or how to see the type IIB cascading dynamics in more details, or even
how to see the $T_N$ building blocks of the Gaiotto models including the recently proposed, tinkertoys \cite{tinkertoys}. All these, and many others, will be 
discussed elsewhere \cite{paseodas}.

\vskip.2in

\centerline{\bf Acknowledgements}

\vskip.1in

\noindent It is a great pleasure to thank Alisha Wissanji for collaboration in \cite{DSW} and the University of Lethbridge for providing a stimulating atmosphere at the
Theory Canada conference where a part of this work was presented by one of us (K. D).
The work of K. D and J. S is supported in parts by NSERC grants.

\bibliographystyle{JHEP}
\bibliography{SWcurve}

\end{document}